\documentclass[usenatbib,usegraphicx]{mn2e}

\title[H{\sc ii} regions and disc surface brightness]{The dependence of H{\sc ii} region properties on global and local surface brightness within galaxy discs}
\author[J. Helmboldt]{J. F. Helmboldt$^{1}$\thanks{E-mail: joe.helmboldt@nrl.navy.mil}, R. A. M. Walterbos$^{2}$, G. D. Bothun$^{3}$, K. O'Neil$^{4}$ and M. S. Oey$^{5}$ \\
$^{1}$Naval Research Laboratory, Code 7213, 4555 Overlook Avenue SW, Washington, DC 20375-5351, USA \\
$^{2}$Department of Astronomy, New Mexico State University, MSC 4500 PO Box 30001 Las Cruces, NM 88003, USA \\
$^{3}$Physics Department, University of Oregon, 1371 East 13th Avenue, Eugene, OR 97403, USA \\
$^{4}$National Radio Astronomy Observatory, P.O. Box 2, Green Bank, WV 24944, USA \\
$^{5}$Astronomy Department, University of Michigan, 830 Dennison Building, Ann Arbor, MI 48109-1042, USA \\
}

\begin{document}

\date{Not yet submitted.}

\pagerange{\pageref{firstpage}--\pageref{lastpage}} \pubyear{2008}

\maketitle

\label{firstpage}

\begin{abstract}
Using B, R, and H$\alpha$ images of roughly equal-sized samples of low surface brightness (LSB) and high surface brightness (HSB) galaxies ($\sim 40$ galaxies apiece), we have explored the dependence of H{\sc ii} region properties on local and global disc surface brightness.  We have done this by constructing co-added H{\sc ii} region luminosity functions (LFs) according to local and central disc surface brightness and fitting Schechter functions to these LFs.  The results show that the shape of the H{\sc ii} region LF within LSB galaxies does not change noticeably as different limiting (i.e., $\mu > \mu_{lim}$) local surface brightness values are used.  However, the LFs for HSB galaxies have larger values of $L_{\ast}$ and are less steep at the faint-end than those of LSB galaxies for limiting B-band local surface brightness values as faint as $\mu_{B,lim} \simeq$23-24.  Both the LFs and the data for individual H{\sc ii} regions show that luminous ($L>10^{39}$ ergs s$^{-1}$) H{\sc ii} regions are much more common within HSB discs than within LSB discs, implying that the newly formed star clusters are also larger.  Taking this into account along with the results of Monte Carlo simulations, the shapes of the LFs imply that the regions within LSB discs and those within the LSB areas of HSB discs are relatively old ($\sim 5$ Myr) while the regions within HSB discs for $\mu_{B} \! \,^{<}_{\sim} 24$ are significantly younger ($< 1$ Myr).  Since the majority of the LSB galaxies do not have noticeable spiral arms and the majority of the HSB galaxies do, this may indicate a transition within HSB discs from spiral arm-driven star formation to a more locally driven, possibly sporadic form of star formation at $\mu_{B} \sim 24$, a transition that does not appear to occur within LSB discs.
\end{abstract}


\section{Introduction}
The completeness of the pioneering optical surveys of galaxies suffered from the inevitable bias toward higher surface brightness produced by the intensity of the night sky.  Digital detectors and data reduction techniques used in modern times have weakened the effect of this bias.  Because of this, the study of so called low surface brightness (LSB) galaxies has become its own sub-field within extragalactic astronomy.  Results from this study have since shown LSB galaxies to be larger at the same luminosity as high surface brightness (HSB) galaxies \citep{deb96} while still obeying the same Tully-Fisher relationship, but only if the entire baryonic content of the galaxies is taken into account \citep{mcg00}.  LSB discs also tend to be less dense \citep{deb98} and have higher gas mass fractions than HSB discs \citep{mcg97}.  In addition, few LSB galaxies have detectable amounts of molecular gas, despite several searches \citep{one00,one03,mat01,one04,mat05}.  LSB galaxies also tend to have lower metallicities \citep[$\sim 1/3$ solar; ][]{mcg94} and are more frequently dwarf and/or irregular galaxies without noticeable spiral arms than HSB galaxies \citep{hel04}.  In addition to this, the H{\sc i } column densities of LSB galaxies tend to be relatively low, at or below the critical limit for star formation \citep{deb96,ken89}.  In light of this, it is not surprising that the star formation rates (SFRs) for LSB galaxies are typically about ten times lower than those of HSB galaxies \citep{mih99,hel04}.  Despite this, LSB discs tend to be relatively blue and some have relatively large H$\alpha$ emission line equivalent widths \citep[see, e.g., ][]{hel04}.\par
These properties of LSB galaxies imply that despite conditions that are not conducive to star formation, many LSB galaxies are actively forming stars.  Because of this, it is possible that they do not form stars in the same way as typical spiral galaxies.  In fact, it has recently been shown that LSB galaxies tend to form fewer high luminosity H{\sc ii} regions and have larger relative amounts of diffuse ionised gas (DIG) than HSB galaxies \citep{hel05,oey07,one07}.  While the larger amounts of DIG may reflect differences in the conditions of the interstellar medium (ISM) within LSB galaxies, the fact that the distribution of H{\sc ii} region luminosities among LSB galaxies is markedly different from that for HSB galaxies points to a possible fundamental difference between how star clusters are formed within these two types of galaxies.  It is remarkable that despite the differences in their star formation histories, LSB galaxies still manage to form roughly the same number of stars at a given circular velocity as HSB galaxies.\par
From recent results, it is not clear if this difference in star formation history results from global or local mechanisms.  In other words, does the lack of spiral arms cause star clusters to form differently throughout a typical LSB disc than throughout an HSB disc, or do star clusters form in a similar way in the LSB regions of HSB discs as they do throughout LSB discs?  To address this question, we have used previously published and newly processed data to compile two roughly equal-sized samples of HSB and LSB galaxies with B, R, and H$\alpha$ images.  We present here a description of the old and new data (\S 2), the results of a detailed analysis of the H{\sc ii} region properties within these two groups of galaxies and how they depend on both local and global properties (\S 3), and Monte Carlo simulations constructed to help interpret these results (\S 4).

\section{Galaxy selection and observations}
To compare properties of H{\sc ii} regions within HSB and LSB galaxies, we have compiled broad-band B and R images and narrow-band H$\alpha$ images of 89 nearby ($V_{r} \;_{\sim}^{<} 6000$ km s$^{-1}$) disc galaxies, 45 HSB galaxies and 44 LSB galaxies.  Based on the so-called Freeman value of $\mu_{0,B} = 21.65 \pm 0.3$ for ``normal'' disc galaxies \citep{fre70}, we have adopted a definition for HSB galaxies of $\mu_{0,B}<21.7$.  Our working definition of an LSB disc galaxy is any galaxy with a B-band central disc surface brightness $>22$ mag arcsec$^{-2}$.  This is one magnitude brighter than what is typically used \citep[see, e.g., ][]{bot97,imp97}, however, it is still more than 1$\sigma$ fainter than the Freeman value.  %We have chosen this limit to (1) provide a larger number of LSB galaxies and (2) to allow for the fact that new observations and surface photometry (see below) of galaxies previously identified as LSB galaxies according to the usual definition have estimated $\mu_{0,B}$ to be as bright as 22.1 for these galaxies.  
In addition, as several authors have demonstrated, the distribution of surface brightness among galaxies is continuous and not bimodal \citep[e.g.,][]{jan00,bla03,hel04}, implying this definition is somewhat arbitrary.  We therefore chose a limit significantly lower than the Freeman value for normal disc galaxies that also yielded a sample of LSB galaxies similar in number to our sample of HSB galaxies.

\subsection{Previously published data}
The images were obtained as part of two separate observing runs in May and October of 2000 conducted at Kitt Peak National Observatory (KPNO) and Cerro Tololo Interamerican Observatory (CTIO), respectively.  The CTIO run consisted of imaging 69 galaxies selected from the H{\sc i} Parkes All Sky Survey \citep[H{\sc i}PASS;][]{bar01} with $V_{r}<2500$ km s$^{-1}$.  All 45 HSB galaxies were obtained from this sample as well as 20 of the LSB galaxies.  The reduction and analysis of the CTIO data was presented in detail by \citet{hel04} and \citet{hel05}, and we will briefly discuss them here.  The images were calibrated using observations of standard stars throughout each night.  For the narrow-band observations during the last two nights of the run, this calibration scheme did not work and a relation was derived using the calibration solutions for the other H$\alpha$ and R-band observations given by the following
\begin{equation}
\Delta m_{H\alpha} = \Delta m_{R} + 2.5 \mbox{log} \left (Ca\frac{\tau_{R}}{\tau_{H\alpha}} \right ) + 0.5
\end{equation}
where $\Delta m$ is the difference between the instrumental magnitude and the actual magnitude, $\tau$ is the exposure time in a particular filter, $a$ is the scale factor each R-band image was multiplied by to be used to subtract the continuum emission from the H$\alpha$ image, and $C$ is a constant empirically determined to be $0.73 \pm 0.15$.  Here, we define the instrumental magnitude to be $-2.5\mbox{log DN}/ \tau + 25$ where DN is the flux measured from the image in ADUs.\par
The 25 mag arcsec$^{-2}$ isophote was then identified on each calibrated B-band image and an ellipse was fit to it, except for one galaxy for which the 26 mag arcsec$^{-2}$ isophote was used because it had $\mu_{0,B}\sim25$.  The position, ellipticity, and position angle of the fitted ellipse was then used to construct elliptical apertures within which the median B and R band surface brightnesses were measured to estimate the surface brightness profile of each galaxy.  The B and R profiles were then simultaneously fit with an exponential disc model for all radii where the surface brightness was between 0.3 and 2.472 mag above the 1$\sigma$ limiting isophote.  This magnitude range was chosen because for an exponential disc, this corresponds to a span of two disc scale lengths.  Each fit was performed with a standard weighted linear least squares routine to obtain the central surface brightness values in each band and a single value for the disc scale length, $h$, as well as error estimates for all three quantities.  The fits were then used to obtain total extrapolated fluxes by integrating each exponential fit from the radius at which the surface brightness is 2.172 mag above the $1\sigma$ limiting isophote to infinity, and then adding this flux to the total flux within that radius measured from the image.\par
\begin{figure}
\includegraphics[scale=0.42]{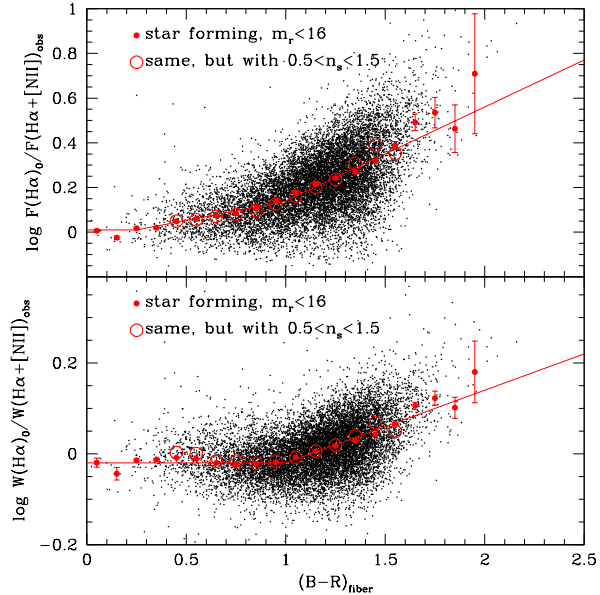}
\caption{The ratio of the H$\alpha$ flux (upper) and equivalent width (lower) to the observed H$\alpha$+[NII] flux for 12,262 star forming SDSS galaxies as measured by \citet{tre04}, including corrections for internal dust extinction according to \citet{cal01}, versus B$-$R colour index as measured from the SDSS spectra (see \S 2.1).  Also plotted are median values within bins of B$-$R and broken power-law fits to the data [not to the median values; see equation(2)].  Median values for galaxies with Sersic indices from \citet{bla03} between 0.5 and 1.5 (i.e., disc-dominated galaxies) are also plotted as large, open red circles.}
\label{hacorr}
\end{figure}

Continuum subtracted H$\alpha$ images were made by scaling the R-band images so that the fluxes of several stars on the R-band and H$\alpha$ images matched, on average, and subtracting the scaled R-band image from the H$\alpha$ image.  The H$\alpha$ filter transmission curve and the galaxy redshift were then used to determine the H$\alpha$+[NII] flux from each calibrated, continuum subtracted H$\alpha$ image.  Using the emission line measurements of \citet{tre04} from Sloan Digital Sky Survey \citep[SDSS;][]{yor00} galaxy spectra, we have derived relationships between B$-$R colour index and the ratios of the H$\alpha$ flux and equivalent width to the observed H$\alpha$+[NII] flux and equivalent width.  The \citet{tre04} measurements included a linear combination of model spectra fit to each galaxy's spectrum so that the effects of stellar absorption would be fully accounted for.  We also corrected both the H$\alpha$ flux and equivalent width measured by \citet{tre04} for internal dust extinction according to \citet{cal01} using the ratio of H$\alpha$ to H$\beta$ emission line flux to compute E(B$-$V) for the ionised gas, assuming H$\alpha$/H$\beta =2.87$ \citep[for T$=10^{4}$ K and case B recombination;][]{ost06}.  Using emission line ratios, we identified 12,262 nearby ($m_{r} < 16$) actively star forming galaxies according to \citet{hel08} from the SDSS data of \citet{tre04} and have measured B$-$R indices from their spectra.  In Fig.\ \ref{hacorr}, we have plotted the ratios of the H$\alpha$ flux and equivalent width to the observed H$\alpha$+[NII] flux and equivalent width versus B$-$R for these galaxies with median values plotted in bins of B$-$R.  Also plotted are broken power-law functions fitted to the data (not the median values) which have the following form

\begin{eqnarray}
X & = & (0.42 \pm 0.066) (B-R) + (-0.28 \pm 0.085) \nonumber \\
  &   & \mbox{if } (B-R) > 0.992, \nonumber \\
  & = & (0.17 \pm 0.11)(B-R) + (-0.032 \pm 0.11) \nonumber \\
  &   & \mbox{if } 0.992 \ge (B-R) > 0.25, \nonumber \\
  & = & (0.011 \pm 0.11) \;\; \mbox{if } (B-R) \leq 0.25. \\
Y & = & (0.16 \pm 0.066) (B-R) + (-0.18 \pm 0.084) \nonumber \\
  &   & \mbox{if } (B-R) > 1, \nonumber \\
  & = & (-0.020 \pm 0.086) \;\; \mbox{if } (B-R) \leq 1.
\end{eqnarray}

\noindent where $X = \mbox{log } F(H\alpha)_{0} / F(H\alpha+[NII])_{obs}$ and $Y = \mbox{log } W(H\alpha)_{0} / W(H\alpha+[NII])_{obs}$.  We must note that the SDSS spectra were obtained using a fibre-fed spectrograph, with each fibre having a 3 arcsec aperture.  This implies that many of the spectra may be bulge-dominated and the above corrections may be more appropriately applied to nuclear regions of galaxies.  To check this, we obtained Sersic indices, $n_{s}$, for the SDSS galaxies from \citet{bla03} and re-computed the median values plotted in Fig.\ \ref{hacorr} for galaxies with $0.5<n_{s}<1.5$, i.e., likely disc-dominated galaxies.  These new median values are plotted in Fig.\ \ref{hacorr} as large, open red circles and are not significantly different from the median values computed for the entire sample of star forming galaxies.  We therefore conclude that it is reasonable to apply the corrections given in equations (2) and (3) to galaxy discs based on their B$-$R colours and not on integrated or nuclear colour indices.\par
We found no significant correlation between the ratios plotted in Fig.\ \ref{hacorr} and either absolute magnitude or surface brightness.  We also note that different empirically derived corrections were used for these data by \citet{hel04} and \citet{hel05}.  However, those previous relations did not include the effects of stellar absorption, and we have therefore concluded that equations (2) and (3) provide better, more extensive corrections for the H$\alpha$+[NII] data.  These new empirical relations with B$-$R colour were used to convert the measured H$\alpha$+[NII] fluxes to extinction corrected H$\alpha$ fluxes using the B$-$R colour of each galaxy's disc computed using the exponential fits to the B and R surface brightness profiles described above (i.e., $(B-R)_{disc} = \mu_{0,B} - \mu_{0,R}$).  We used the  $(B-R)_{disc}$ colours because (1) the vast majority of the star formation within these galaxies occurs within their discs, and (2) the exponential fits yield B$-$R colour estimates for the discs that are reasonably free from the effects of any substantial small scale colour fluctuations and bulge or bar contamination.\par%  The scaled R-band images where then used with these fluxes to produce integrated H$\alpha$ emission line equivalent widths.\par
To measure the properties of the H{\sc ii} regions, the HII{\it phot} algorithm developed by \citet{thi00} was used.  The algorithm is designed to identify H{\sc ii} regions by smoothing the H$\alpha$ images with kernels of different sizes and fitting models to significant peaks in the smoothed images.  The boundary of each H{\sc ii} region is then grown until a limiting emission measure gradient is reached (a limit of 1.5 EM pc$^{-1}$ was used), or until the boundary reaches the boundary of another region.  Following this, surface fits to designated background pixels are used to estimate the amount of diffuse emission along the lines of sight passing through each H{\sc ii} region \citep[see ][for more detailed descriptions]{thi00,hel05}.
\subsection{Newly processed data}
To supplement the imaging data for LSB galaxies available from the CTIO observing run, a sample of 32 galaxies previously identified as LSB disc galaxies was observed at KPNO.  The 32 targets were culled from the literature \citep{sch83,bot85,kra86,elm87,tul88,sch88,sch90,one97} and were chosen to represent the wide variety of known LSB disc galaxies, ranging from dwarf (disc scale lengths $\sim$500 pc) to giant (scale lengths $\sim$10 kpc) LSB galaxies.  However, because of cloudy conditions during the observing run, we did not perform standard star observations and the usual method of calibrating the images could not be used.  In the intervening years, the completion of the bulk of the imaging for the SDSS has provided a means to calibrate the B and R band images of 24 of the LSB galaxies observed.  We have done this by measuring the fluxes of several ($\sim$6-10) stars on the KPNO B and R images which have measured $g$ and $r$ magnitudes from SDSS images.  The $g$ and $r$ magnitudes were converted to B and R magnitudes using the conversions given by \citet{smi02}, and the mean difference between the raw KPNO fluxes and these magnitudes was used to calibrate each image.  The typical uncertainty in these calibration solutions is about 0.06 mag and 0.02 mag for the B and R bands, respectively.\par
%In general, the difference between the instrumental and calibrated magnitudes increases with airmass, as expected (see Fig.\ \ref{ext}).  However, there is a large amount of scatter and the trend does not follow the expected relation for the typical atmospheric extinction for KPNO ($k_{B}=0.46$ and $k_{R}=0.19$), possibly owing to the non-photometric conditions of the observing run.  For one galaxy, UGC 9380, the difference between the instrumental and calibrated magnitudes is unusually large (2.76 mag; see Fig.\ \ref{ext}).  However, we note that there is only a $0.5 \pm 0.3$ mag difference between our calibrated B-band total magnitude (see below) and that from the RC3 catalogue \citep{dev91}.  This, along with log entries attesting to the weather conditions at the time of the observation, lead us to believe that the large value of $m_{B,inst}-m_{B}$ was due to clouds and that our calibration scheme has suitably accounted for this.  In addition to UGC 9380, there are 11 other 
There are 12 galaxies from our KPNO sample with B-band magnitudes from the RC3 catalogue.  As Fig.\ \ref{photcheck} demonstrates, our total B-band magnitudes (see below) agree with the RC3 values to within about $1\sigma$ on average and that the difference between the two is not systematic and does not depend on the brightness of the galaxy.  Note that the $1\sigma$ uncertainties in our total B-band magnitudes are combinations of the calibration errors, the photon errors, and the errors in the parameters from the fits to the surface brightness profiles used to compute fluxes integrated to radii of $\infty$ (see below).\par

%\begin{figure}
%\includegraphics[scale=0.42]{airmass_ext.ps}
%\caption{The difference between the instrumental (see \S 2.1) and calibrated magnitudes for the LSB galaxies observed at KPNO for the B (upper) and R (lower) bands versus airmass.  In each panel, the dashed line is a weighted linear least squares fit to the data and the dotted line is the result predicted by the mean KPNO extinction curve.  For UGC 9380, the B-band observation was exceptionally non-photometric ($m_{inst}-m=2.76$), but the measured B-band magnitude agrees with the value given in the RC3 catalogue within roughly $1\sigma$.}
%\label{ext}
%\end{figure}

\begin{figure}
\includegraphics[scale=0.42]{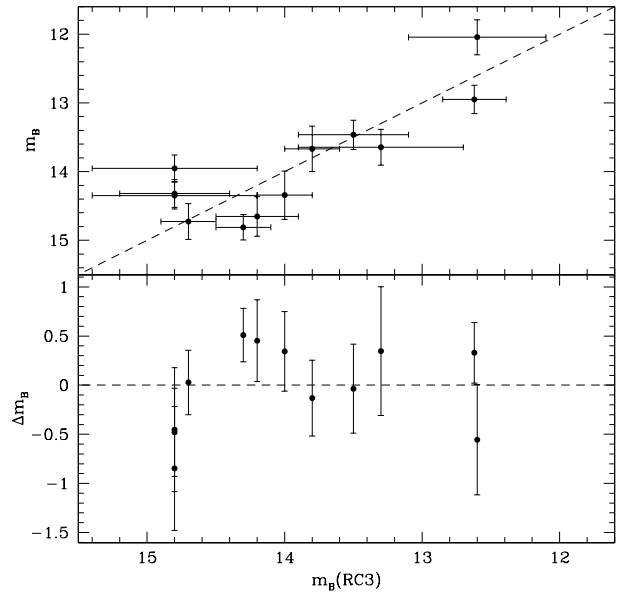}
\caption{In the upper panel, the total extrapolated B-band magnitude (see \S 2.1) for the 12 LSB galaxies observed at KPNO that are in the RC3 catalogue versus the RC3 B-band magnitudes.  The difference between the B-band magnitude measured using our KPNO images and the RC3 magnitude is plotted versus the RC3 magnitude in the lower panel.  In each panel, the dashed line represents the case where the magnitudes are the same and is not a fit to the data.}
\label{photcheck}
\end{figure}

\begin{figure}
\includegraphics[scale=0.42]{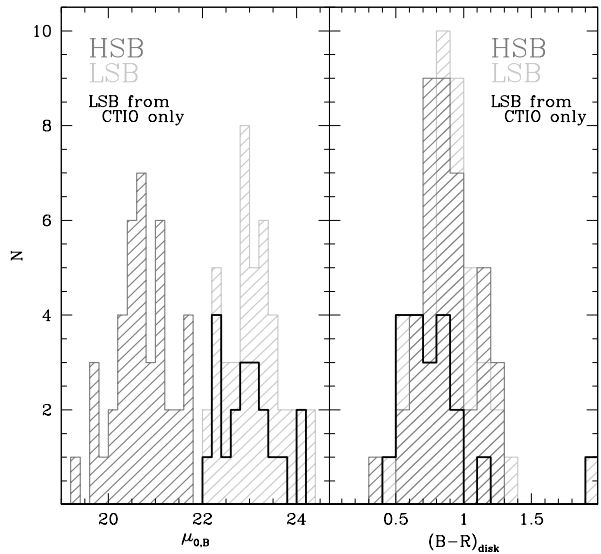}
\caption{The B-band central surface brightness (left) and B$-$R disc colours (right; see \S 2.1) distributions for the HSB (grey) and LSB (light grey) galaxies.  The distributions for the LSB galaxies from the CTIO data only (black histograms) are also included.}
\label{mudist}
\end{figure}

\begin{figure*}
\includegraphics[scale=0.9]{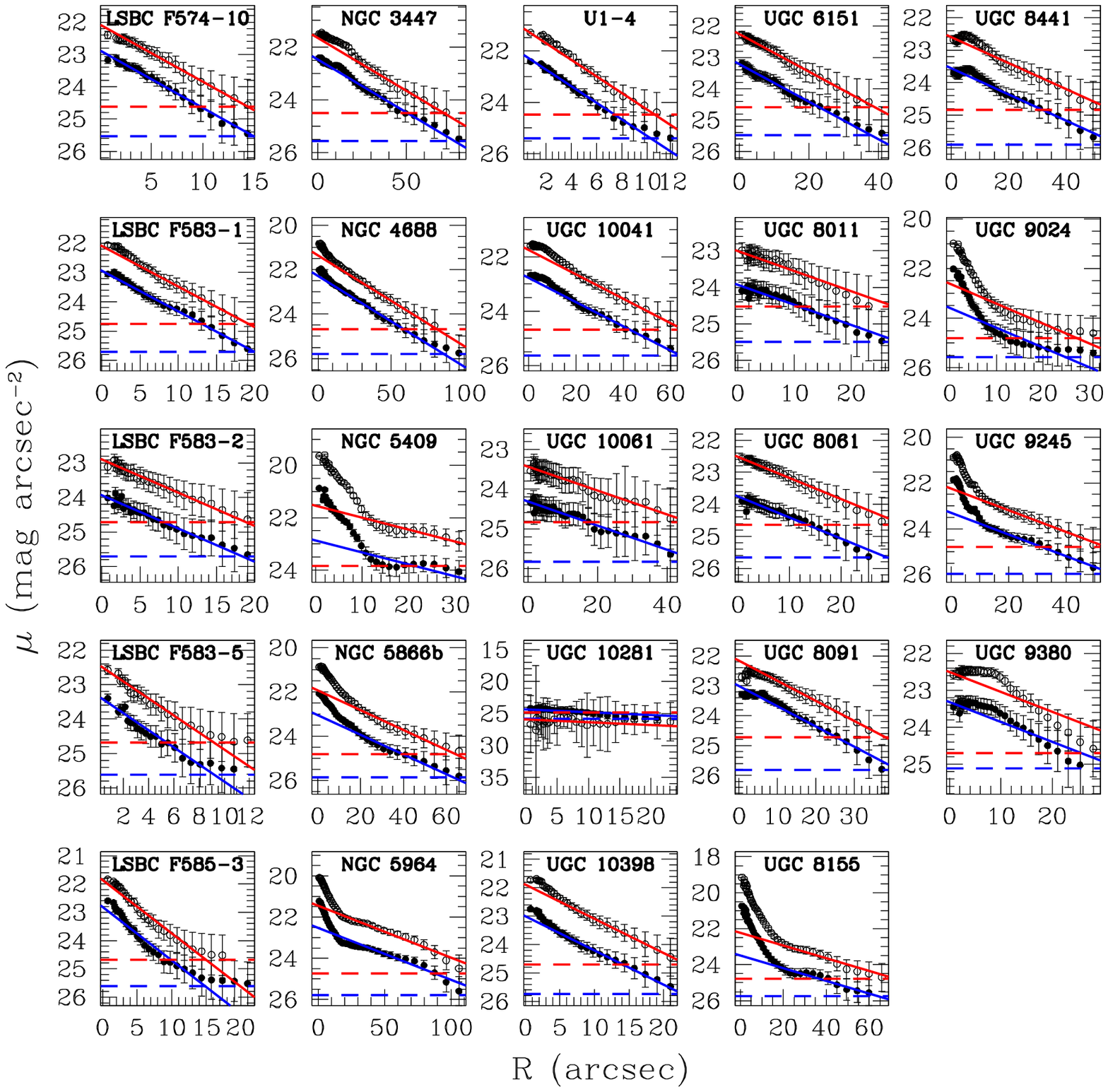}
\caption{The R (open red points) and B (solid blue points) band surface brightness profiles for the LSB galaxies observed at KPNO; no offsets have been applied to the data.  The best fitting exponential profiles (see \S 2.1) are plotted in red and blue for the R and B band profiles, respectively.  The $1\sigma$ limiting isophote is plotted as a horizontal dashed line for each band.}
\label{profs}
\end{figure*}

For consistency, the surface photometry for these new images was performed in the same way as the CTIO images, with one important difference.  Since many of these galaxies have relatively low central surface brightnesses, the B-band 25 mag arcsec$^{-2}$ isophote was not ideal to use to characterise the shape of each galaxy if it could be identified at all.  Instead, a different isophote was used for each galaxy, determined by eye through trial and error by varying the isophote level in 0.5 mag steps.  The isophotes used are listed in Table \ref{kpnodat} and range from 26 to 27 mag arcsec$^{-2}$.  We have also listed in Table \ref{kpnodat} the ellipticity and position angle of this isophote along with the exponential disc scale length, the B and R central surface brightnesses, and the total extrapolated B and R magnitudes, all determined in the same way as the CTIO data \citep[see above and ][]{hel04}.  The surface brightness profiles for all 24 galaxies are displayed in Fig.\ \ref{profs} along with the exponential disc fits.  The $\mu_{0,B}$ distributions plotted in the left panel of Fig.\ \ref{mudist} show that the inclusion of the newly processed KPNO data not only doubles the number of LSB galaxies, it also provides a much better sampling of the $23.5 < \mu_{0,B} < 24.5$ region.  In the right panel of Fig.\ \ref{mudist}, we have plotted the distributions of $(B-R)_{disc}$.  These distributions show that the inclusion of the new KPNO data has also introduced a significant number of relatively red LSB discs that supplement the relatively blue LSB discs found among the CTIO data.  These redder discs have made the $(B-R)_{disc}$ distributions for the HSB and LSB galaxies essentially the same.\par

\begin{table*}
\centering
\caption{Properties of LSB Galaxies Observed at KPNO}
\begin{scriptsize}
\begin{tabular}{lccccccccccc@{$\;$}c@{$\;$}c}
\hline
 & R.A. (J2000) & Dec. (J2000) & $V_{r}$ &  & PA &  & $h$ &  &  &  &  & log $F(H\alpha)$ & log $W(H\alpha)$ \\
Name & (h m s) & ($\circ$ $'$ $''$) & (km s$^{-1}$) & $e$ & ($^{\circ}$) & $\mu_{iso}$ & ($''$) & $\mu_{0,B}$ & $\mu_{0,R}$ & $m_{B}$ & $m_{R}$ & (ergs/s/cm$^{2}$) & ($\mbox{\AA}$) \\
(1) & (2) & (3) & (4) & (5) & (6) & (7) & (8) & (9) & (10) & (11) & (12) & (13) & (14) \\
\hline
  NGC    3447 &   10 53 26.8 &    $+$16 46 44.0 & 1067 & 0.32 & $-$73.0 & 26.0 & 27.2 & 22.5 & 21.6 & 13.1 & 12.3 &   $-$11.9 &     1.71 \\ 
 UGC    6151 &  11 05 56.3 &  $+$19 49 31.0 & 1335 & 0.18 &  88.0 & 27.0 & 18.2 & 23.2 & 22.3 & 14.8 & 13.9 &   $-$13.0 &     1.33 \\ 
        U1-4 &  11 38 25.8 &  $+$17 05 03.1 & 3452 & 0.38 &  69.9 & 26.0 &  3.4 & 22.1 & 21.1 & 17.3 & 16.3 &   $-$13.7 &     1.11 \\ 
 NGC    4688 &  12 47 46.5 &  $+$04 20 09.9 &  986 & 0.16 &  39.7 & 26.5 & 26.6 & 22.3 & 21.4 & 12.9 & 12.1 &   $-$11.9 &     1.49 \\ 
LSBC F574-10 &  12 50 31.4 &  $+$17 28 18.7 &  844 & 0.72 & $-$74.5 & 26.0 &  6.1 & 22.9 & 22.1 & 16.8 & 16.1 &   $-$14.0 &     0.94 \\ 
 UGC    8011 &  12 52 21.1 &  $+$21 37 46.2 &  776 & 0.08 & $-$12.2 & 27.0 & 19.7 & 23.9 & 23.0 & 15.3 & 14.4 &   $-$13.2 &     1.43 \\ 
 UGC    8061 &  12 56 44.0 &  $+$11 55 55.3 &  562 & 0.17 & $-$86.0 & 27.0 & 16.6 & 23.8 & 22.5 & 15.5 & 14.3 &   $-$13.5 &     1.22 \\ 
 UGC    8091 & 12 58 40.4 & $+$14 13 02.9 &  214 & 0.39 & $-$74.1 & 26.5 & 16.2 & 23.0 & 22.2 & 14.7 & 13.9 &   $-$12.0 &     2.16 \\ 
 UGC    8155 &  13 03 14.7 &  $+$07 48 07.9 & 2925 & 0.24 & $-$86.2 & 27.0 & 30.8 & 23.5 & 22.2 & 13.6 & 12.4 &   $-$12.5 &     1.07 \\ 
 UGC    8441 &  13 25 29.1 &  $+$57 49 20.0 & 1519 & 0.17 &   1.6 & 27.0 & 27.2 & 23.6 & 22.6 & 14.3 & 13.4 &   $-$12.8 &     1.31 \\ 
 NGC    5409 &  14 01 46.1 &  $+$09 29 25.0 & 6259 & 0.01 &  12.1 & 26.0 & 23.8 & 22.9 & 21.5 & 13.7 & 12.3 & $\cdots$ & $\cdots$ \\ 
 UGC    9024 & 14 06 40.6 & $+$22 04 12.1 & 2323 & 0.16 & $-$26.8 & 27.0 & 13.5 & 23.6 & 22.6 & 15.6 & 14.6 &   $-$13.3 &     1.26 \\ 
 UGC    9245 & 14 25 26.9 & $+$56 19 12.6 & 1885 & 0.30 & $-$79.3 & 27.0 & 23.0 & 23.3 & 22.2 & 14.3 & 13.3 &   $-$12.7 &     1.22 \\ 
 UGC    9380 &  14 34 39.2 &  $+$04 15 46.0 & 1693 & 0.32 & $-$66.5 & 27.0 & 20.2 & 23.3 & 22.5 & 14.7 & 13.9 &   $-$12.8 &     1.50 \\ 
 NGC   5866B & 15 12 07.2 & $+$55 47 06.3 &  841 & 0.45 & $-$82.5 & 26.5 & 24.3 & 23.0 & 21.9 & 14.0 & 12.8 &   $-$12.7 &     1.26 \\ 
LSBC  F583-2 &   15 37 00.5 &    $+$20 47 42.0 & 1719 & 0.36 &  38.8 & 27.0 & 11.3 & 23.9 & 22.9 & 16.4 & 15.5 &   $-$13.7 &     1.50 \\ 
 NGC    5964 &  15 37 36.3 &  $+$05 58 24.3 & 1447 & 0.09 & $-$27.4 & 26.0 & 43.2 & 22.5 & 21.4 & 12.0 & 11.0 &   $-$12.1 &     1.01 \\ 
LSBC  F583-5 &  15 45 43.9 &  $+$17 18 41.0 & 3261 & 0.34 &  28.4 & 27.0 &  4.4 & 23.3 & 22.4 & 18.0 & 17.1 &   $-$13.8 &     1.69 \\ 
 UGC   10041 &  15 49 01.4 &  $+$05 11 19.0 & 2171 & 0.62 &  89.0 & 26.5 & 24.4 & 22.8 & 21.8 & 13.5 & 12.5 &   $-$12.3 &     1.36 \\ 
 UGC   10061 &  15 51 15.1 &  $+$16 19 46.4 & 2080 & 0.34 &  88.6 & 27.0 & 36.1 & 24.3 & 23.4 & 14.3 & 13.5 &   $-$13.0 &     1.59 \\ 
LSBC  F583-1 &  15 57 27.5 &  $+$20 39 58.0 & 2264 & 0.49 &  86.2 & 26.5 &  7.8 & 22.9 & 22.1 & 16.3 & 15.5 &   $-$13.5 &     1.51 \\ 
 UGC   10281 &  16 13 20.6 &  $+$17 11 34.3 & 1084 & 0.04 &  69.4 & 27.0 & 24.9 & 24.4 & 25.9 & 15.1 & 16.8 &   $-$15.6 &     0.61 \\ 
LSBC  F585-3 &  16 21 23.8 &  $+$20 51 56.0 & 3100 & 0.22 &  42.8 & 26.5 &  5.8 & 22.8 & 21.8 & 16.7 & 15.9 &   $-$13.3 &     1.60 \\ 
 UGC   10398 &  16 28 10.5 &  $+$17 38 23.9 & 4529 & 0.19 & $-$16.2 & 26.5 &  9.0 & 23.0 & 21.9 & 16.0 & 15.0 &   $-$13.7 &     1.16 \\
\hline
\end{tabular}
\end{scriptsize}
\caption{Col. (4): Radial velocity.  Col. (5): Isophotal ellipticity.  Col. (6): Isophotal position angle (from north through east).  Col. (7): B-band surface brightness of isophote used for Col. (5) and (6).  Col. (8): Exponential disc scale-length.  Col. (9): Disc central B-band surface brightness.  Col. (10): Disc central R-band surface brightness.  Col. (11): Total extrapolated B-band magnitude (see \S 2), corrected for foreground dust extinction.  Col. (12): Total extrapolated R-band magnitude (see \S 2), corrected for foreground dust extinction.  Col. (13): Log of the total H$\alpha$ flux (see \S 2).  Col. (14): Log of the total H$\alpha$ emission line equivalent width (see \S 2).}
\label{kpnodat}
\end{table*}

\begin{figure*}
\includegraphics[scale=0.9]{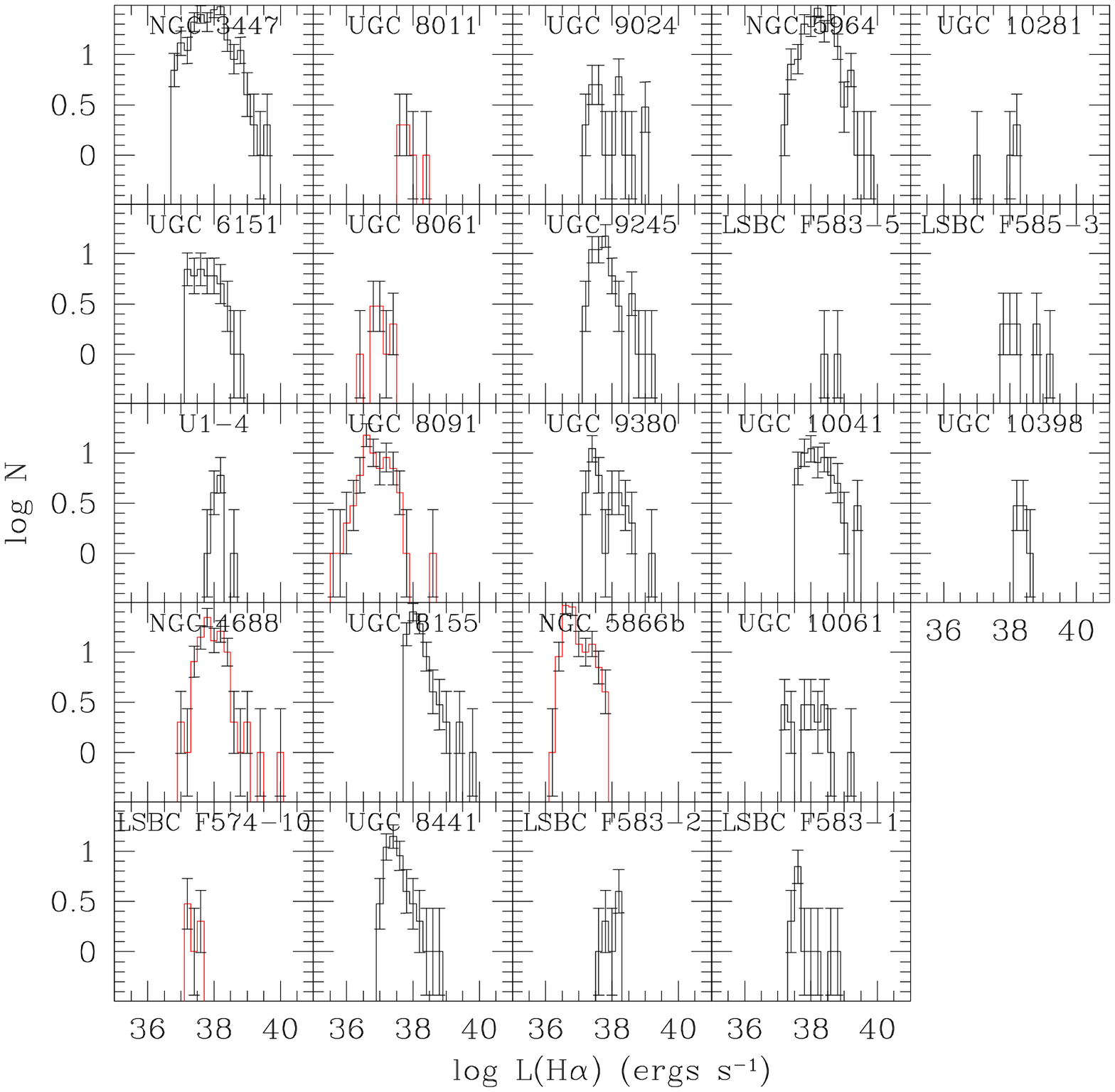}
\caption{The H{\sc ii} region luminosity functions (LFs) for the LSB galaxies observed at KPNO.  LFs for galaxies with radial velocities $<$1,000 km s$^{-1}$ for which the distance estimates are somewhat unreliable are highlighted in red.}
\label{lfs}
\end{figure*}

Without standard star observations or spectra, we have no direct means of calibrating the H$\alpha$ images obtained for these galaxies.  However, for all but five of the galaxies observed a KPNO, the same set of 30 $\mbox{\AA}$ wide filters, kindly provided by Richard Rand, was employed as was used for most of the CTIO observations.  For the remaining galaxies, two broader H$\alpha$ filters provided at KPNO were used which are similar to those available at CTIO which were employed for 21 of the 69 galaxies observed there.  Because the same or similar H$\alpha$ filters were used and since we were again using scaled versions of the R-band images to subtract the continuum emission from the H$\alpha$ images, it was possible to use the derived R-band calibrations with equation (1) to calibrate the H$\alpha$ images.  We have done this assuming the mean value of $C=0.73$ determined using the CTIO observations and have also included the corrections for internal dust extinction, [NII] contamination, and stellar absorption derived from the \citet{tre04} data given in equations (2) and (3), again using the disc B$-$R disc colours computed using the exponential fits to the surface brightness profiles (see \S 2.1).  We note that the uncertainty in the value of $C$ corresponds to an uncertainty of about 0.09 dex, which is comparable or less than the uncertainties in the corrections given by equations (2) and (3).  The resulting calibrated continuum subtracted H$\alpha$ images were then used to measure the total H$\alpha$ flux and equivalent width for each galaxy, which is listed in Table \ref{kpnodat}.  We have also used HII{\it phot} to identify and perform photometry on H{\sc ii} regions for each of these images using the same settings as were used for the CTIO images.  We note that for one of the LSB galaxies observed at KPNO, NGC 5409, no H$\alpha$ image was obtained.  The H{\sc ii} region luminosity functions (LFs) of the 23 galaxies with H$\alpha$ images were computed using the HII{\it phot} results, assuming H$_{0}=$70 km s$^{-1}$ Mpc$^{-1}$, and are displayed in Fig.\ \ref{lfs}.  LFs for galaxies with radial velocities $<1000$ km s$^{-1}$ for which the distances may be relatively uncertain are highlighted in red.  The H{\sc ii} region properties and LFs of the galaxies observed at CTIO are presented in \citet{hel05}.

%\section{H{\sc ii} Regions and DIG: LSB versus HSB}
\section{H{\sc ii} regions}
\subsection{Global properties}
\begin{figure}
\includegraphics[scale=0.42]{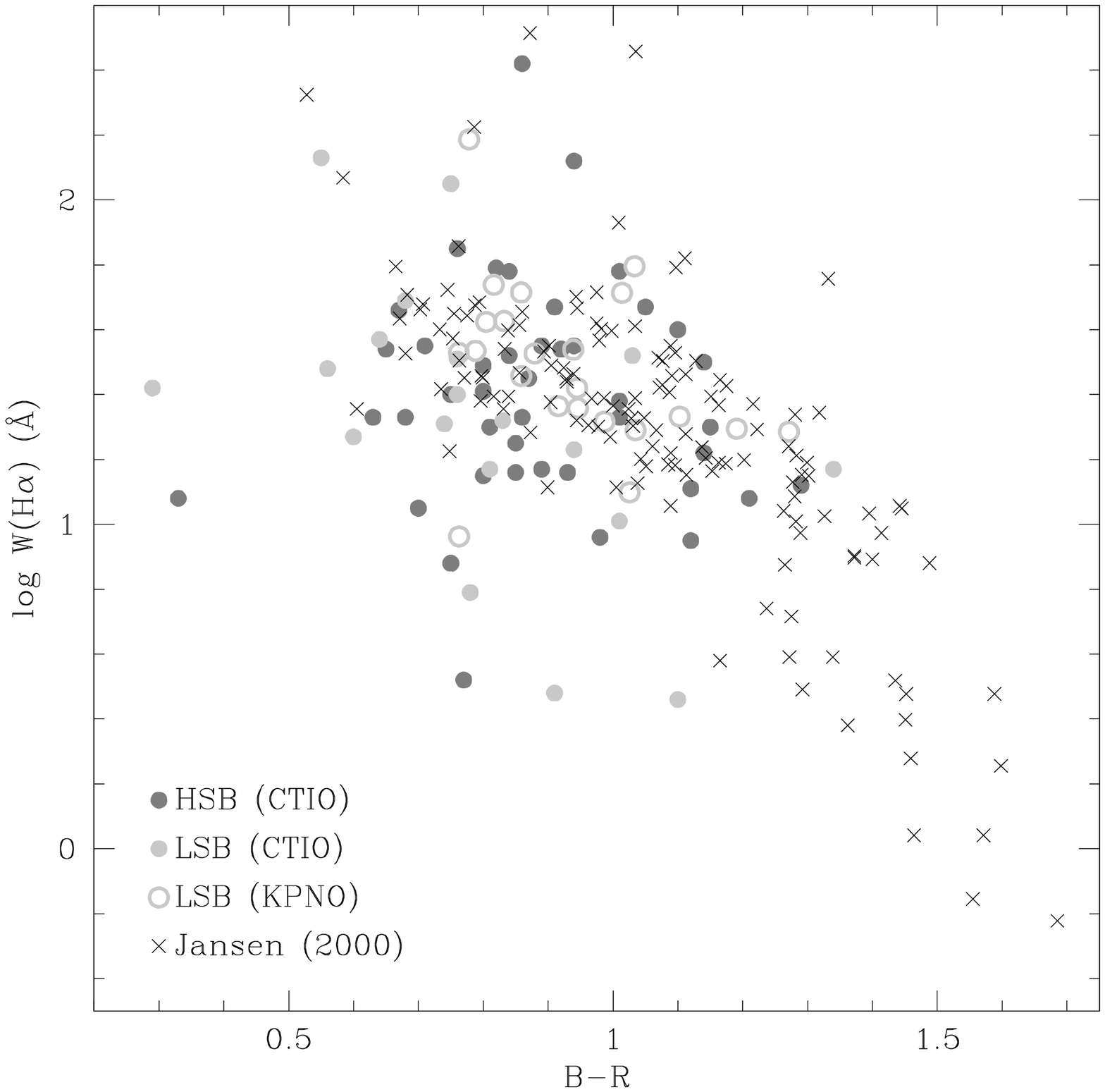}
\caption{The total H$\alpha$ emission line equivalent width versus the B$-$R colour index for HSB (grey) and LSB (light grey) galaxies; LSB galaxies observed at KPNO are plotted as open points.  For comparison, the galaxies from the Nearby Field Galaxy Survey of Jansen (2000) are plotted with $\times$ symbols.}
\label{sfh}
\end{figure}

With B, R, and H$\alpha$ images for 45 HSB and 43 LSB galaxies, we are in a good position to explore the differences in H{\sc ii} region properties between the two classes of galaxies.  In Fig.\ \ref{sfh}, we have plotted the total H$\alpha$ emission line equivalent width as a function of B$-$R index for the HSB and LSB galaxies as well as similar data from the Nearby Field Galaxy Survey \citep[NFGS;][]{jan00}, which includes a wide range of morphological types.  From this plot, it can be seen that the amounts of star formation relative to the stellar continua within both types of galaxies are similar.  Both are bluer on average than the NFGS galaxies.  For the most part, both the LSB and HSB galaxies have similar H$\alpha$ equivalent widths at the same colour indices as the NFGS galaxies.\par
\begin{figure}
\includegraphics[scale=0.42]{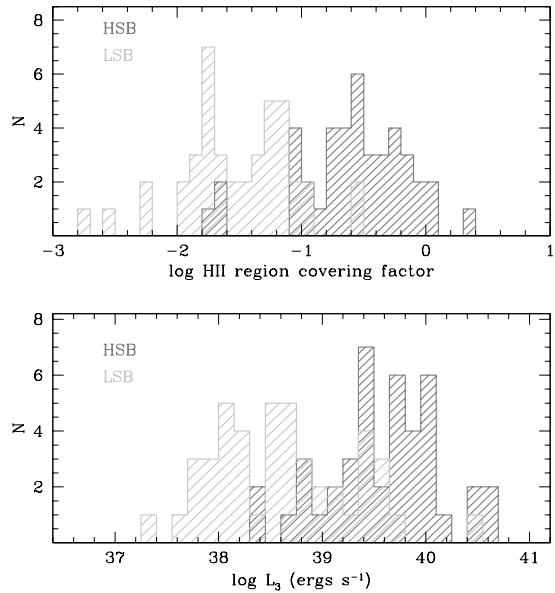}
\caption{Distributions for the H{\sc ii} region covering factor (see \S 3.1; upper) and the mean H$\alpha$ luminosity of the three brightest H{\sc ii} regions (lower) for the HSB (grey) and LSB (light grey) galaxies.}
\label{hiidig}
\end{figure}

However, we start to see distinct differences between HSB and LSB galaxies when we look at H{\sc ii} region properties.  In Fig.\ \ref{hiidig}, we have plotted the distributions for two properties (1) the H{\sc ii} region covering factor, the ratio of the total area occupied by H{\sc ii} regions taken from the HII{\it phot} results to the total disc area taken to be $9\pi h^{2}$, where $h$ is the disc scale length, %(i.e., the area of a circle with a radius of $3h$), %(2) the diffuse fraction, $f_{d}$, the ratio of the H$\alpha$ flux attributed to the DIG using the HII{\it phot} results to the total H$\alpha$ flux, 
and (2) the mean luminosity of the three brightest H{\sc ii} regions, $L_{3}$.  We note that for the surface brightness profiles presented in \citet{hel04} and in Fig.\ \ref{profs}, the radii are the geometric mean radii of the elliptical apertures used, i.e., $R= \sqrt{ab}$.  Therefore, the assumed total disc area of $9 \pi h^{2}$ is the area of an ellipse with a semi-major axis of $3 h (b/a)^{-1/2}$, where $b/a$ is the axis ratio of the elliptical apertures used.  No attempt has been made to de-project the estimated total disc area or the area occupied by H{\sc ii} regions.  However, we have found no significant correlation between $b/a$ and either the H{\sc ii} region covering factor or the average number of regions per square kiloparsec with Spearman rank correlation coefficients of -0.08 and -0.18, respectively.  Additionally, the mean axis ratios for the LSB and HSB samples are similar ($<\!\! b/a \!\!>=0.65 \mbox{ and } 0.60$, respectively).  It is therefore unlikely that this has had a significant effect on the results presented here.\par
As Fig.\ \ref{hiidig} demonstrates, the H{\sc ii} region covering factor is significantly lower on average for LSB galaxies than it is for HSB galaxies, and the same is true for $L_{3}$.  %The diffuse fraction is higher for LSB galaxies on average and near unity in some cases.  
These results are generally consistent with what has been found previously \citep{gre98,hel05,oey07,one07}.\par

To compare the typical distributions of H{\sc ii} region luminosities, we have constructed co-added H{\sc ii} region LFs for both the LSB and HSB galaxies.  First, we must note that we only have radial velocity distances for our galaxies which implies that those with smaller radial velocities will have larger distance uncertainties.  These large errors may artificially alter the shape of any co-added LF, and we therefore must exclude such galaxies from this process.  \citet{hel05} determined that the optimum radial velocity cut-off for a co-added LF with bins of 0.2 dex in width is $V_{r}<840$ km s$^{-1}$.  For all co-added LFs discussed in this paper (see \S 3.2 as well), we have excluded galaxies with $V_{r}<840$ km s$^{-1}$ and have used the same bin width as \citet{hel05}.  There are 40 of our HSB galaxies and 37 of our LSB galaxies that have $V_{r}>840$ km s$^{-1}$ .\par
After applying the radial velocity cut, the next step to co-adding the galaxies' LFs was to determine a limiting H{\sc ii} region luminosity for each galaxy by fitting a power-law to the H{\sc ii} region luminosity as a function of the signal-to-noise ratio, S/N, of the integrated flux of the region as determined by HII{\it phot}.  The limiting luminosity was then taken to be the luminosity given by the power-law fit for S/N$=5$, plus the rms deviation of the data about the power-law fit.  We then constructed an H{\sc ii} region LF for each galaxy above this limiting luminosity.  Each LF was then divided by the area of the disc, $9 \pi h^{2}$, to make sure that larger galaxies with more H{\sc ii} regions would not dominate the shape of the co-added LF.  As noted above, this estimate of the total disc area has not been de-projected to the face-on disc area.  However, as also noted, there is no significant correlation between the isophotal axis ratio, $b/a$, and the average number of H{\sc ii} regions per square kiloparsec, and the mean values of $b/a$ for the LSB and HSB samples are similar.\par
For each galaxy group, the average number of regions per square kiloparsec was computed for each luminosity bin by averaging over all galaxies with limiting H{\sc ii} region luminosities larger than the upper boundary of the bin.  Uncertainties for each LF were computed using jackknife resampling.  Specifically, the galaxies were put in groups of four and the LF was computed 11 different times, each time excluding one group of four, and the standard deviation for each luminosity bin over these 11 computations was taken to be the uncertainty.  For instances where the shot noise ($\sqrt{N_{tot}}$, where $N_{tot}$ is the total number of regions among all co-added galaxies) was larger than the jackknife value, the shot noise was adopted as the uncertainty.\par
\begin{figure}
\includegraphics[scale=0.42]{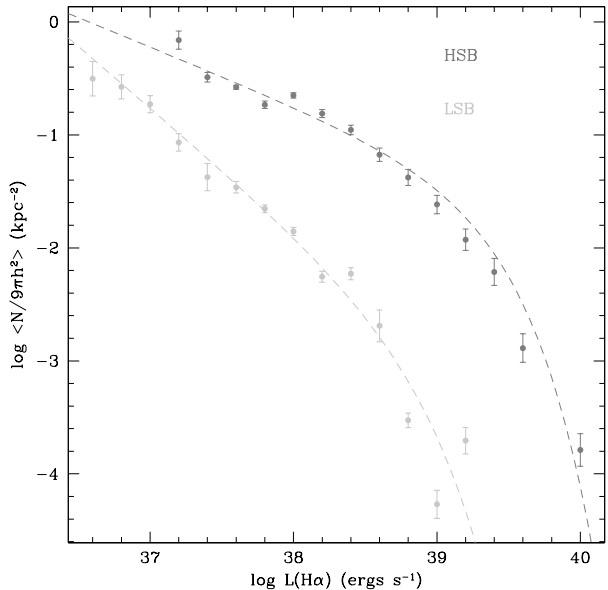}
\caption{Co-added H{\sc ii} region luminosity functions (see \S 3.1) for the HSB (grey) and LSB (light grey) galaxies with radial velocities $>$840 km s$^{-1}$ (see \S 3.1).  %The LF for regions from HSB galaxies with local B-band surface brightness $>24$ is also included, plotted with open grey points.  
Schechter functions fit to the LFs are also plotted as dashed lines.}
\label{coadd}
\end{figure}

The HSB and LSB galaxy co-added LFs are plotted with these uncertainties in Fig.\ \ref{coadd}.  We can see that while the LSB LF is well approximated by a single power-law over more than two decades in H$\alpha$ luminosity, the HSB LF shows a significant amount of curvature.  This is similar to what was found by \citet{hel05}.  To quantify this, we have fit to the LFs the following function
\begin{equation}
\frac{dN}{dL} = \phi_{\ast} \left ( \frac{L}{L_{\ast}} \right )^{\alpha} \mbox{exp} \left ( \frac{-L}{L_{\ast}} \right )
\end{equation}
commonly referred to as a Schechter function where $\alpha$ is the faint-end slope and $L_{\ast}$ characterises the amount of curvature at the bright-end.  The Schechter function was chosen because it is basically a truncated power-law which is qualitatively what one would expect for an aged population of star clusters with a power-law distribution of initial masses (see the Monte Carlo simulations described in \S 4).  The fitted functions are also plotted in Fig.\ \ref{coadd}.  The values of $\alpha$ and $L_{\ast}$ from these fits are -2.08 and $5.79 \times 10^{38}$ ergs s$^{-1}$ for the LSB LF and -1.52 and $1.86 \times 10^{39}$ ergs s$^{-1}$ for the HSB LF.  We note that this difference in LF shape is unlikely to be artificially caused by the application of the corrections given in equation (2) since these corrections are solely functions of $(B-R)_{disc}$, and Fig.\ \ref{mudist} demonstrates that the $(B-R)_{disc}$ distributions for the LSB and HSB galaxies are essentially the same.  We must also acknowledge that the shape of the LF, particularly for HSB galaxies for which the H{\sc ii} region covering factor is larger, may be artificially flattened by blending of H{\sc ii} regions \citep[see][]{sco01}.  We will address this issue in more detail in \S 3.3.

\subsection{Local properties}
\begin{figure}
\includegraphics[scale=0.42]{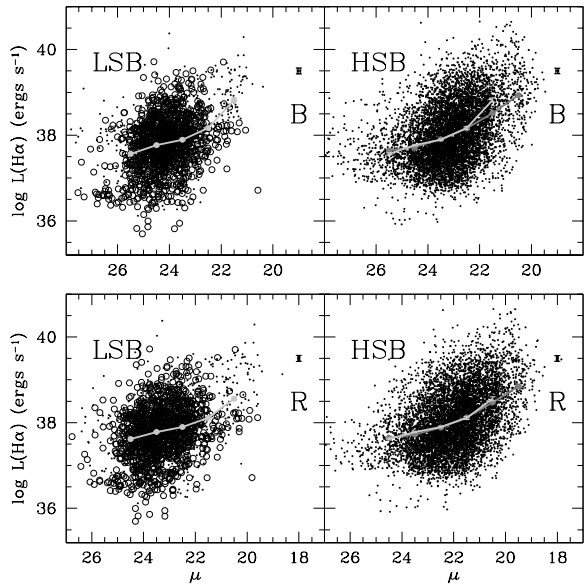}
\caption{The H$\alpha$ luminosity of H{\sc ii} regions from LSB (left) and HSB (right) galaxies as a function of local surface brightness in the B (upper) and R (lower) bands.  In the left panels, data for H{\sc ii} regions within the LSB galaxies observed at KPNO are plotted as open circles.  Mean values are plotted within surface brightness bins for both LSB (light grey) and HSB (grey) galaxies.}
\label{loclum}
\end{figure}

While the results presented in Fig.\ \ref{hiidig} and \ref{coadd} mostly reinforce earlier results, a question still remains regarding whether they arise from differences in the global properties of HSB and LSB galaxies, or from local conditions within the LSB and HSB regions of all discs.  While results previously presented by \citet{gre98} suggest that these properties vary with local surface brightness within HSB discs, we now have the ability to explore whether or not that variation is seen in LSB discs and if it is indeed similar to what has been seen for HSB discs.  We have done this by measuring the local B and R band surface brightness in the area near each H{\sc ii} region using a circular aperture with a radius of $2 \sqrt{n_{p}} / \pi$ where $n_{p}$ is the number of pixels occupied by the H{\sc ii} region on the image.  We have defined the sizes of the apertures in this way rather than using an aperture with a fixed physical size to ensure that each aperture was small enough to represent the local conditions around each region while being large enough that the region itself did not dominate the local flux.  %In Fig.\ \ref{locdfrac}, we have plotted the local diffuse fraction versus local B and R surface brightness separately for regions within LSB and HSB galaxies.  Median values of $f_{d}$ within local surface brightness bins show that while the local $f_{d}$ dispersion is large, there is a tendency for lower surface brightness areas to have higher diffuse fractions.  These median values also show that this trend is virtually the same for LSB and HSB galaxies for both the B and R bands.  
In Fig.\ \ref{loclum}, we have plotted the H{\sc ii} region luminosity versus local surface brightness.  For both LSB and HSB galaxies, there appears to be a trend for H{\sc ii} regions to be more luminous in higher surface brightness areas.  The fact that there are few low luminosity H{\sc ii} regions in higher surface brightness areas may be a selection effect as they are harder to detect in such areas.  However, the paucity of highly luminous H{\sc ii} regions in more LSB areas cannot be the result of such a bias, implying that while this possible selection effect may influence the observed trend, it cannot be the cause of it.  In Fig.\ \ref{loclum}, for $\mu_{B}<23$, the trend for LSB galaxies becomes somewhat steeper than that for HSB galaxies.  However, this is heavily influenced by a few LSB galaxies which contain some HSB regions and this difference is not seen in the R-band data.  In both the B and R data, the trend appears to become steeper for $\mu_{B}<22.5$ and $\mu_{R}<21.5$.\par
\begin{figure}
\includegraphics[scale=0.42]{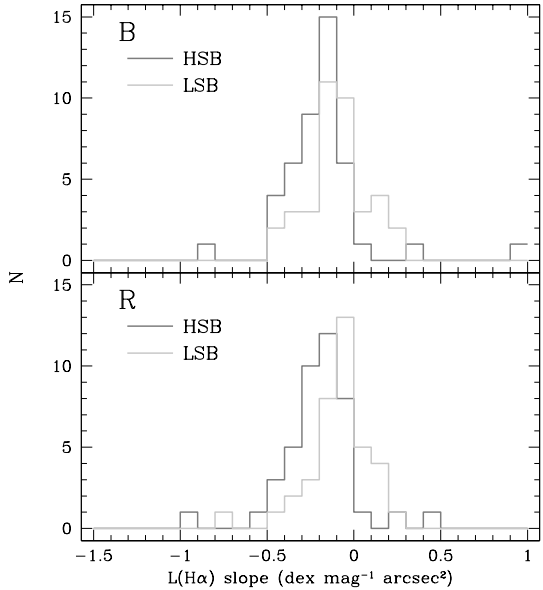}
\caption{The distributions of the gradients of the log of the H{\sc ii} region H$\alpha$ luminosity (see \S 3.2) as a function of local surface brightness in the B (upper) and R (lower) bands for H{\sc ii} regions within LSB (light grey) and HSB (grey) galaxies.}
\label{grads}
\end{figure}

To further explore and constrain the results shown in Fig.\ %\ref{locdfrac} and 
\ref{loclum}, we have fit a line to %$f_{d}$ and 
log $L(H\alpha)$ as a function of local surface brightness for each galaxy to obtain a gradient.  The distributions for these gradients using both the B and R data are plotted in Fig.\ \ref{grads} separately for LSB and HSB galaxies.  %The $f_{d}$ gradient distributions are similar for both LSB and HSB galaxies with median and peak values which are $>0$, confirming the results seen in Fig.\ \ref{locdfrac}.  
On average, the log $L(H\alpha)$ gradients for LSB galaxies appear to be about 0.1 dex mag$^{-1}$ arcsec$^{2}$ less steep than those for HSB galaxies with both distributions peaking at values $<0$.  This is likely a reflection of the fact that the LSB galaxies have a substantially higher fraction of regions with $\mu_{B}>22.5$ and $\mu_{R}>21.5$, where the trends seen in Fig.\ \ref{loclum} are less steep, than HSB galaxies.  We note that the gradient distributions for both the B and R bands peak around $-0.15$ to $-0.05$ dex mag$^{-1}$ arcsec$^{2}$, which is similar to the slopes for the median values plotted in Fig.\ \ref{loclum}.%  This is less steep than the trend between surface brightness at the half-light radius and $L_{3}$ found by \citet{hel05} using the CTIO data which has a slope of about $-0.25$ dex mag$^{-1}$ arcsec$^{2}$.
\par

\begin{figure}
\includegraphics[scale=0.42]{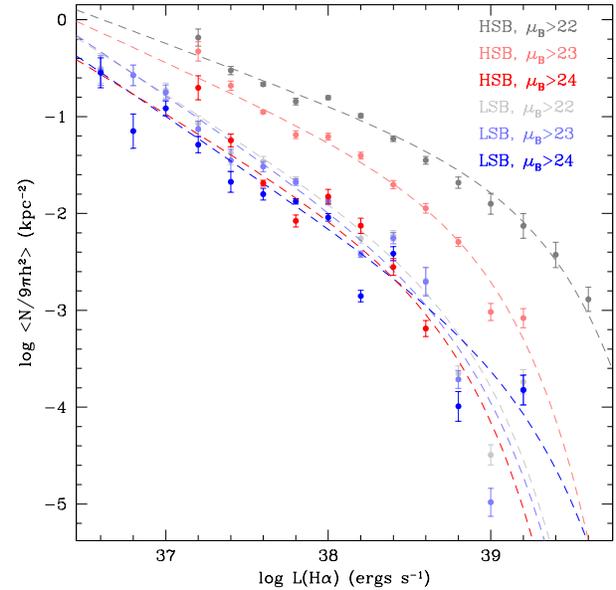}
\caption{Co-added LFs for H{\sc ii} regions from HSB galaxies (grey to red) and from LSB galaxies (light grey to blue) using different limiting local B-band surface brightness values (see panel).}
\label{loclf}
\end{figure}

\begin{figure}
\includegraphics[scale=0.42]{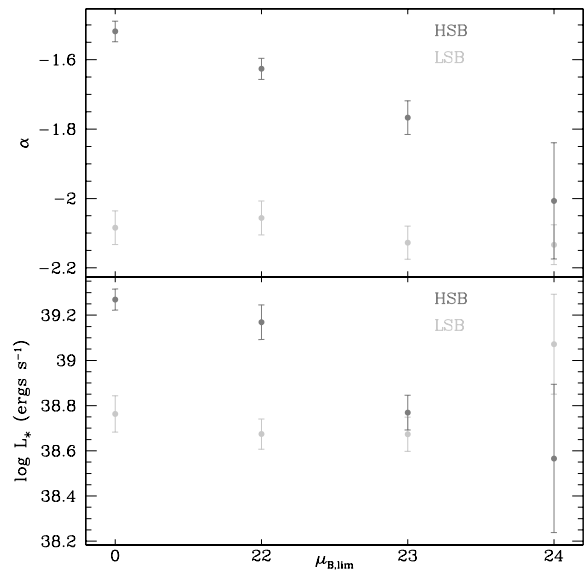}
\caption{For the LFs displayed in Fig.\ \ref{coadd} and  \ref{loclf}, the faint-end slope, $\alpha$ (upper), and the value of $L_{\ast}$ (lower) for the Schechter functions fit to the LFs as a function of limiting local B-band surface brightness.  The parameters from the fits to the LFs using all H{\sc ii} regions are plotted for $\mu_{B,lim} = 0$.  As in previous figures, the results for HSB galaxies are plotted in dark grey and the results for LSB galaxies in light grey.}
\label{lfshape}
\end{figure}

%The fact that the correlation between H{\sc ii} region luminosity and local surface brightness is noticeably weaker than the trend between $L_{3}$ and galaxy surface brightness suggests that it is possible that the shape of the H{\sc ii} region LF may be more significantly affected by global rather than local conditions.
The results illustrated by Fig.\ \ref{loclum} and \ref{grads} imply that for a given local surface brightness, the typical H{\sc ii} region luminosity for LSB galaxies is likely similar to that for HSB galaxies.  This implies that local surface brightness/density strongly affects star cluster formation within galaxy discs, as suggested by the toy model of \citet{one98}.  However, this does not mean that global disc properties do not have a significant effect on the detailed properties and formation histories of star clusters.  For instance, at the same local surface brightness, the shapes of the distributions of H{\sc ii} region luminosities are not necessarily the same for LSB and HSB galaxies, and any difference in shape may point to the influence of global properties on cluster formation.  To explore this further, we have constructed co-added LFs for H{\sc ii} regions based on local B-band surface brightness separately for HSB and LSB galaxies.  We did this by making three co-added LFs separately for LSB and HSB galaxies using three different limiting local B-band surface brightness values, $\mu_{B,lim} =$22, 23, and 24 mag arcsec$^{-2}$.  In other words, each co-added LF was remade using only H{\sc ii} regions with local $\mu_{B} > \mu_{B,lim}$.  The six resulting LFs are plotted in Fig.\ \ref{loclf} along with Schechter function fits.  From these plots, it can be seen that the HSB LFs have a noticeably larger amount of curvature and a faint-end with a shallower slope than the LSB LFs for $\mu_{B,lim} < 24$.  This is better illustrated in Fig.\ \ref{lfshape} where we have plotted the parameters of the Schechter functions fitted to the LFs in both Fig.\ \ref{coadd} and \ref{loclf} as functions of $\mu_{B,lim}$.  While the faint-end slope, $\alpha$, remains roughly constant with $\mu_{B,lim}$ for LSB galaxies, it appears to become steeper with decreasing limiting surface brightness for HSB galaxies.  At $\mu_{B,lim} = 24$, $\alpha$ appears to be essentially the same for both LSB and HSB galaxies.  A similar pattern can be seen for the $L_{\ast}$ parameter, except that the values for LSB and HSB galaxies appear to be in good agreement for $\mu_{B,lim}$ as bright as about 23 mag arcsec$^{-2}$.

%\begin{figure}
%\includegraphics[scale=0.42]{coaddLF_mu_LSB.ps}
%\caption{Co-added LFs for H{\sc ii} regions from LSB galaxies grouped by local B-band surface brightness (see \S 3.2) in 1 mag arcsec$^{-2}$ increments.  The LF in the upper left panel is for all H{\sc ii} regions within LSB galaxies with radial velocities $>$840 km s$^{-1}$ (see \S 3.1).  The best fitting Schechter function (see equation (2)) for each LF is plotted as a dashed line.}
%\label{loclf}
%\end{figure}

%\begin{figure}
%\includegraphics[scale=0.42]{coaddLF_mu_HSB.ps}
%\caption{Co-added LFs for H{\sc ii} regions from HSB galaxies grouped by local B-band surface brightness (see \S 3.2) in 1 mag arcsec$^{-2}$ increments.  The LF in the upper left panel is for all H{\sc ii} regions within HSB galaxies with radial velocities $>$840 km s$^{-1}$ (see \S 3.1).  The best fitting Schechter function (see equation (2)) for each LF is plotted as a dashed line.}
%\label{loclf}
%\end{figure}

\begin{figure*}
\includegraphics[scale=0.8]{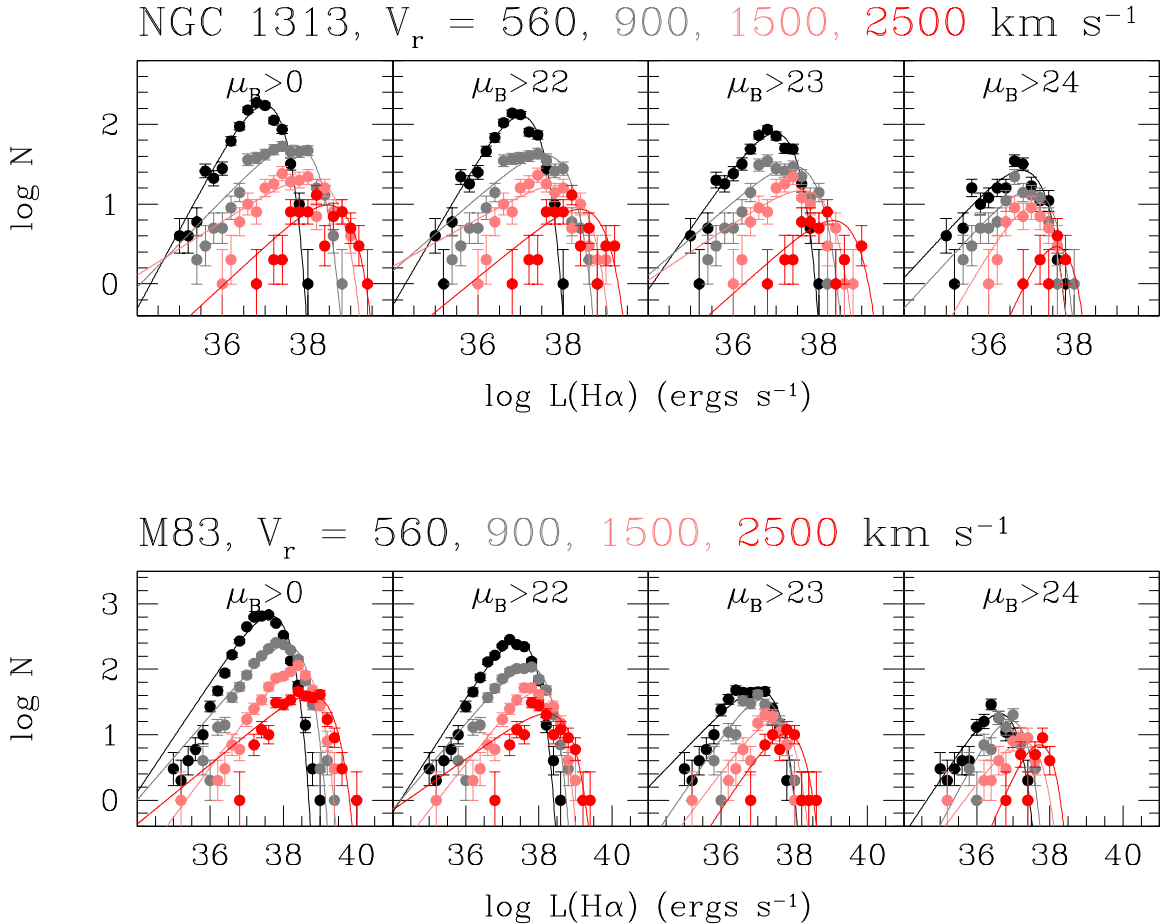}
\caption{For NGC 1313 (upper panels) and M83 (lower panels), the H{\sc ii} region LFs for four different values of $\mu_{B,lim}$ (see panels) at four different simulated radial velocities (see \S 3.3).  The solid lines are Schechter functions fit to the data.}
\label{blendlf}
\end{figure*}

\subsection{Blending}
Since we have used ground-based images of galaxies with radial velocities $>840$ km s$^{-1}$ to examine H{\sc ii} region properties in disc galaxies, we cannot ignore the influence blending has on the results displayed in Fig.\ \ref{loclf} and \ref{lfshape}.  Blending may be particularly important when considering the difference between the shapes of the H{\sc ii} region LFs in HSB and LSB areas within discs.  This is evidenced by the fact that in Fig.\ \ref{loclf}, the HSB galaxy LFs have a significantly larger number of regions per square kiloparsec than the LSB galaxy LFs for $\mu_{lim,B}<23$, implying that blending may be much more important for these LFs.  Since the shapes of the LFs for HSB and LSB galaxies are roughly similar for $\mu_{B,lim}\!\,^{>}_{\sim} 23$, it is possible that this difference in shape is purely the result of the LFs of regions from HSB areas within discs being artificially biassing toward higher luminosities by blending.\par
The effect of blending can be demonstrated by smoothing images of more nearby HSB galaxies and reanalysing them with HII{\it phot}.  For this, we have selected two HSB galaxies that were imaged during the October 2000 CTIO run, M83 and NGC 1313.  These two galaxies were chosen because they are relatively nearby ($V_{r}=$516 and 457 km s$^{-1}$, respectively) and have $\sim$1,000 H{\sc ii} regions identified by HII{\it phot} so that LFs can be constructed for each galaxy for different values of $\mu_{lim,B}$ with significant numbers of regions.  In addition to this, M83 and NGC 1313 have two distinct morphologies; M83 is a ``grand design'' spiral galaxy (type Sb) and NGC 1313 is an SBd galaxy with a somewhat disturbed morphology, especially in H$\alpha$ emission.  To simulate the effect of blending at different radial velocities, we smoothed the B, R, and H$\alpha$ images of each galaxy with the appropriate Gaussian kernel to simulate the size of the seeing disc measured from stars on the narrow-band image at 10 radial velocites, 560, 600, 700, 800, 900, 1100, 1300, 1500, 2000, and 2500 km s$^{-1}$.  The images were then re-gridded so that the angular size of the pixels relative to that of the seeing disc remained constant.  Following this, Gaussian noise was added to each image so that the noise in the smoothed image was similar to that of the original image.  HII{\it phot} was then used to identify the H{\sc ii} regions and measure their properties on each of these images.\par
\begin{figure}
\includegraphics[scale=0.42]{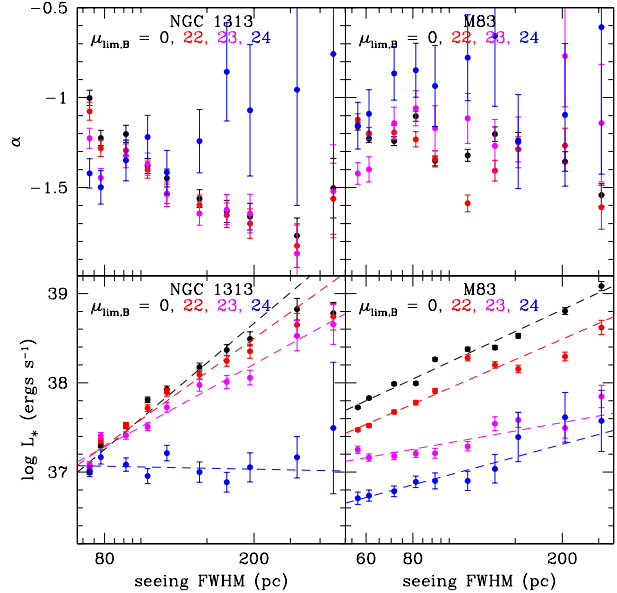}
\caption{For NGC 1313 and M83, the parameters $\alpha$ and $L_{\ast}$ derived from the Schechter functions fit to the LFs measured at 10 simulated radial velocities (560, 600, 700, 800, 900, 1100, 1300, 1500, 2000, and 2500 km s$^{-1}$; see \S 3.3) for four different values of $\mu_{B,lim}$ versus the physical size of the seeing disc.  The dashed lines in the bottom panels are weighted, linear least squares fits to the data plotted to illustrate the convergence (or lack thereof) of the trends between $L_{\ast}$ and the spatial resolution among the different values of $\mu_{B,lim}$.}
\label{lffit}
\end{figure}

Following the HII{\it phot} analysis, we constructed H{\sc ii} region LFs for $\mu_{B,lim}=$0, 22, 23, and 24 for each galaxy at each radial velocity.  The resulting LFs for $V_{r}=$560, 900, 1500, and 2500 km s$^{-1}$ are plotted in Fig.\ \ref{blendlf} with Schechter function fits.  In general, as $V_{r}$ increases, the number of detected regions decreases and the LFs become more biassed toward higher luminosities.  To better demonstrate this, we have plotted the values of $\alpha$ and $L_{\ast}$ from the Schechter function fits as a function of the physical size of the seeing disc in Fig.\ \ref{lffit}.  For both galaxies, $\alpha$ is not significantly different for different values of $\mu_{B,lim}$ for $\mu_{B,lim} \leq 23$.  For $\mu_{B,lim}=24$, $\alpha$ is slightly higher.  For M83, $\alpha$ does not appear to change significantly with radial velocity while for NGC 1313, $\alpha$ decreases with increasing $V_{r}$ for $\mu_{B,lim} \leq 23$.  We therefore find no evidence from these two examples that blending has caused $\alpha$ to be larger in higher surface brightness areas of HSB discs as seen in Fig.\ \ref{lfshape}.\par
For both NGC 1313 and M83, $L_{\ast}$ increases with $V_{r}$ for $\mu_{B,lim} \leq 23$ and for all values of $\mu_{B,lim}$ for M83; $L_{\ast}$ remains roughly constant for NGC 1313 for $\mu_{B,lim} = 24$.  For NGC 1313, the value of $L_{\ast}$ appears to differ somewhat among the different values of $\mu_{B,lim}$ with the value for $\mu_{B,lim} = 24$ being substantially lower at nearly all radial velocities.  However, the values of $L_{\ast}$ appear to converge at the lowest radial velocities.  In contrast, the differences among the values of log $L_{\ast}$ for different limiting surface brightnesses for M83 are roughly constant with $V_{r}$.  Linear fits to these data (see Fig.\ \ref{lffit}) suggest that the values of $L_{\ast}$ may converge for relatively small (between $\sim 2$ and 10 pc) spatial resolutions.  However, this requires one to extrapolated the trends in the lower right panel of Fig.\ \ref{lffit} roughly one order of magnitude lower than the smallest spatial resolution at $V_{r}=560$ km s$^{-1}$, implying that there is no evidence that blending alone may be able to reproduce the trend seen in the lower panel of Fig.\ \ref{lfshape} for M83.  From these two examples, it is clear that the values of $L_{\ast}$ for the higher surface brightness areas of HSB galaxies are likely inflated by the effects of blending.  However, it is also clear that while for some galaxies that are similar to NGC 1313, the trend between $L_{\ast}$ and $\mu_{B,lim}$ may be entirely caused by blending, for other, more typical spiral galaxies like M83, the trend is somewhat unaffected as blending appears to increase $L_{\ast}$ by roughly the same factor for all values of $\mu_{B,lim}$.  It is therefore likely that the trend seen in the lower panel of Fig.\ \ref{lfshape} has a real physical origin and is not entirely the result of blending.\par
We note that the Schechter fits were not restricted to similar luminosity ranges as those fit to the LFs in Fig.\ \ref{loclf} because the radial velocity distances for these two galaxies are highly uncertain.  This has possibly led to the inclusion of more faint H{\sc ii} regions and sharper turn-overs at the faint end.  We therefore note that the results presented in Fig.\ \ref{lffit} most likely do not reflect the magnitude of the effect of blending on the trend shown in Fig.\ \ref{lfshape}.  However, the results are still useful for testing whether or not blending is chiefly responsible for the trends seen for HSB galaxies in Fig.\ \ref{lfshape}.  From the results shown in Fig\ \ref{lffit}, it appears that blending is likely no the sole cause of these trends.

\section{Interpretation and discussion}
As the results presented in Fig.\ \ref{loclf} and \ref{lfshape} have shown, the shape of the H{\sc ii} region LF appears to be substantially affected by global processes.  In particular, both the faint-end slope and $L_{\ast}$ seem to be roughly independent of local surface brightness for LSB galaxies, while both parameters seem to decrease as the limiting local B-band surface brightness is made fainter for HSB galaxies.  %while the faint-end slope of the LF appears to change with local surface brightness, the amount of curvature at the bright-end seems to be roughly independent of local surface brightness and much more prominent for HSB galaxies than for LSB galaxies.  
In \S 3.3, we demonstrated with two examples that the effects of blending have probably inflated the values of $L_{\ast}$, possibly to a larger degree for higher surface brightness areas.  However, we also demonstrated that it is likely that blending is not the sole cause of the results shown in the lower panel of Fig.\ \ref{lfshape}.  To analyse the physical origins of the difference in LF shapes between LSB and HSB galaxies, we have constructed Monte Carlo simulations of H{\sc ii} region LFs patterned after those developed by \citet{oey98} which are similar to those used by \citet{von90}.  We note that \citet{oey98} explored different assumptions about the time dependence of the H$\alpha$ luminosity due to stars of different masses and about the formation history of star clusters.  However, they found that assuming a constant H$\alpha$ luminosity over the main sequence lifetime of each star which abruptly goes to zero and an instantaneous burst where all clusters form at once was able to reproduce the shapes of typical H{\sc ii} region LFs.  We have therefore adopted these two assumptions as well as the same power-laws for the distribution of the number of stars per cluster (slope of $\beta = -2$), the relationships between both H$\alpha$ luminosity and main sequence lifetime and stellar mass (slopes of $\delta=1.5$ and $d=-0.7$, respectively), and the initial mass function (IMF) of \citet{sal55}.  We have also assumed the same upper and lower mass limits for the IMF of 17 and 100 M$_{\odot}$ and an H$\alpha$ luminosity and main sequence lifetime for 100 M$_{\odot}$ stars of $3 \times 10^{38}$ ergs s$^{-1}$ and 2.8 Myr.  We note that these power-law relations break down below the adopted lower mass limit for the IMF, which artificially introduces a turn-over at log $L(H\alpha) \sim 37.7$ and we do not consider the Monte Carlo LFs below this luminosity.  Note that since this is a non-physical limit, it was not applied to the fitting of Schechter functions to the observed LFs.\par

%\begin{figure}
%\includegraphics[scale=0.42]{alpha_lstar.ps}
%\caption{For the LFs displayed in Fig.\ \ref{loclf} and \ref{loclf}, the faint-end slope, $\alpha$ (upper), and the value of $L_{\ast}$ (lower) for the Schechter functions fit to the LFs as a function of local B-band surface brightness.  As in previous figures, the results for HSB galaxies are plotted in dark grey and the results for LSB galaxies in light grey.  In both panels, the horizontal dashed lines represent the values for the Schechter function fits to all H{\sc ii} regions (see the upper left panels in  Fig.\ \ref{loclf} and \ref{loclf}).  For both parameters, values with large ($>$1) errors and values obtained from fits using $\leq 6$ data points were considered unreliable and were excluded from this plot.}
%\label{lfshape}
%\end{figure}

Beyond the assumptions we have made, the two physical factors that determine the shapes of the LFs are the age (from birth) of the clusters and the maximum number of stars per cluster, $N_{\ast , max}$.  To explore how these two parameters affect the faint-end slope and amount of curvature at the bright-end, we have constructed 121 LFs with 1,000 clusters apiece for 11 ages ranging from 0 to 6 Myr and 11 values of $N_{\ast , max}$ ranging from 10 to 1,000.  Each simulated LF used the same set of random numbers to assign a number of stars to each of the 1,000 clusters from the assumed power-law distribution, truncated at $N_{\ast}=N_{\ast , max}$.  We note that while ages up to 6 Myr may seem too large for H{\sc ii} regions, the instantaneous burst model of \citet{bru03} estimates that at an age of 6 Myr, the flux of ionising photons output by a cluster of stars would have dropped by a factor of roughly 30, but would be far from negligible.\par
An additional factor that must also be considered is the uncertainty in the H$\alpha$ luminosity which varies from one galaxy to another and can be rather large because of the statistical corrections used for internal dust extinction, $[$NII$]$ contamination, and stellar absorption [see equation (2)].  Also, for galaxies for which the R-band calibration and equation (1) were used to calibrate the H$\alpha$ flux, the uncertainty in the H$\alpha$ luminosity is even larger given the uncertainty in the $C$ parameter in equation (1).  This technique was used for a larger fraction of the LSB galaxies and may therefore contribute to any differences between the shapes of the LSB and HSB LFs.  To incorporate this into our Monte Carlo-simulated LFs, we first computed an estimate of the full uncertainty in the H$\alpha$ calibration for each galaxy.  This computation included the uncertainties in the parameters given in equation (2).  It also included the combination of the uncertainty in the $C$ parameter in equation (1) and the uncertainty in the R-band calibration, or, in the case of the majority of the CTIO observations, the uncertainty in the calibration solution for the H$\alpha$ filters derived from observations of spectrophotometric standard stars.  Finally, a radial velocity uncertainty of 94 km s$^{-1}$ was assumed for each galaxy which is based on the uncertainty in the correction for in-fall into the Virgo cluster \citep{hel05}.  The resulting uncertainties in log $L(H\alpha)$ range from about 0.1 to 0.2 dex.  %, and generally dwarf the photometric uncertainties in even the faintest H{\sc ii} regions (S/N$\simeq 5$, or uncertainty of $\sim 0.087$ dex).  
For each group of galaxies, HSB and LSB, the fraction of the total number of H{\sc ii} regions per square kiloparsec among all galaxies in the group was computed for each galaxy.  This was then used to assign the uncertainty of that galaxy to the same fraction of H{\sc ii} regions in the Monte Carlo simulations.  This uncertainty was applied to the log of the H$\alpha$ luminosity of each simulated region by adding to it a factor of $\sigma g$ where $g$ is a random number drawn from a unit normal distribution and $\sigma$ is the assigned uncertainty.  Since all of the H{\sc ii} regions within a particular galaxy have the same H$\alpha$ calibration uncertainty, the same value of $g$ was used for all simulated regions with the same assigned value of $\sigma$.  Each of the 121 simulated LFs was reconstructed twice, once using the uncertainties for the LSB group, and once using those of the HSB group, resulting in 363 simulated LFs in all.\par
\begin{figure*}
\includegraphics[scale=0.9]{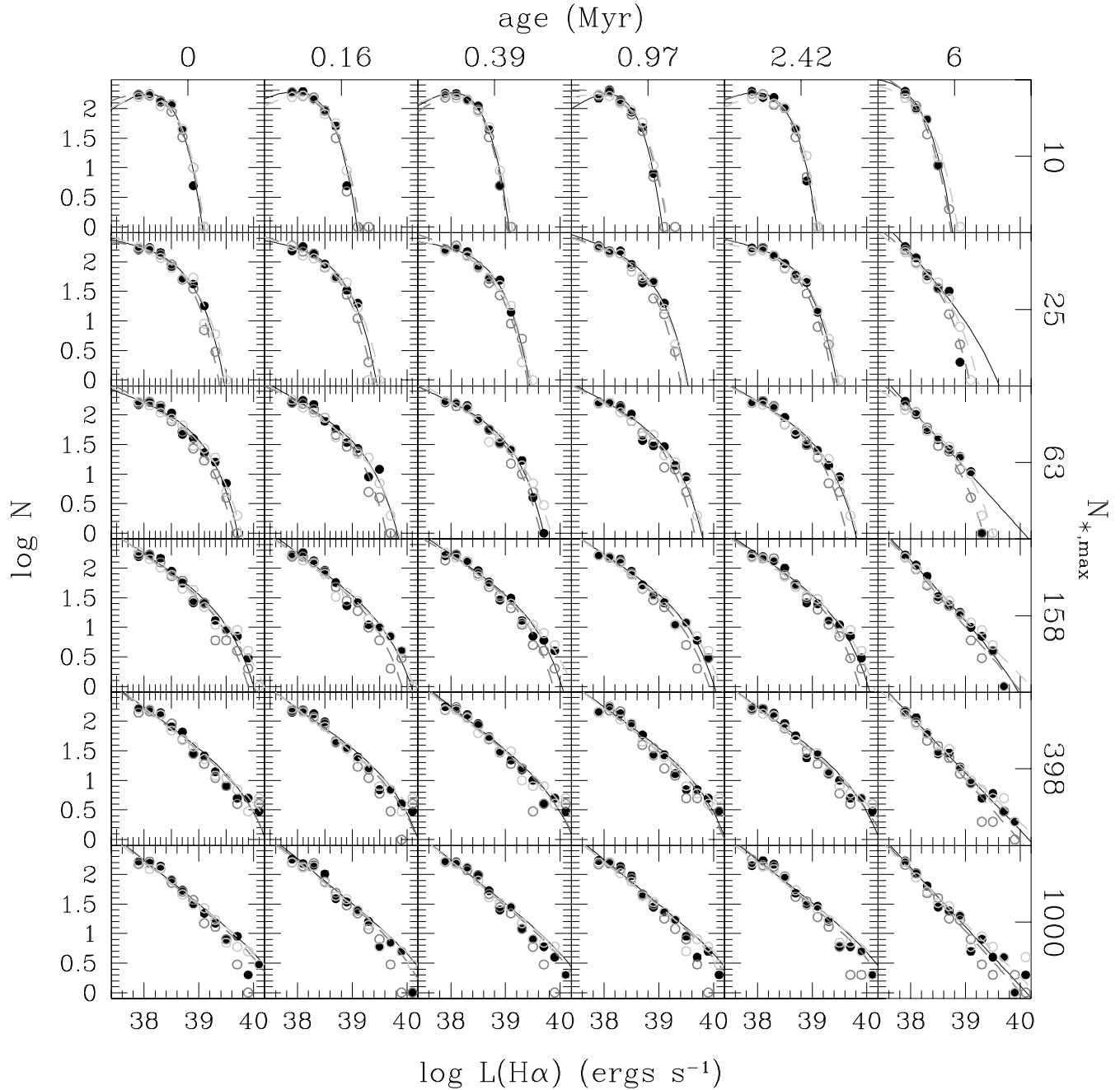}
\caption{A sampling of the Monte Carlo simulated H{\sc ii} region LFs for six star cluster ages (marked above the panels) and six values of $N_{\ast , max}$, the maximum number of stars per cluster (marked to the right of the panels).  The simulations used 1,000 stellar clusters apiece and were pattered after those of \citet[][see \S 4.2]{oey98}.  The simulated LFs that do not include modelled $L(H\alpha)$ uncertainties are plotted in black.  Those that included modelled uncertainties based on the estimated H$\alpha$ calibration uncertainties (see \S 4) for LSB and HSB galaxies are plotted in light grey and grey, respectively.  The plots shown here also include Schechter function fits [see equation (2)].}
\label{mclfs}
\end{figure*}

Schechter functions were fit to all of the simulated LFs, including those with and without the modelled effects of calibration uncertainty.  A sampling of the simulated LFs for combinations of six values of star cluster age and six values of $N_{\ast , max}$ are displayed in Fig.\ \ref{mclfs} with their Schechter fits.  Those without the modelled effects of H$\alpha$ calibration uncertainty are plotted in black; those with modelled uncertainties based on the HSB and LSB groups are plotted in grey and light grey, respectively.  These plots show that for a fixed value of either $N_{\ast , max}$ or age, the amount of curvature at the bright-end decreases as the other parameter increases.  However, it does appear that $N_{\ast , max}$ has a much stronger effect on the amount of curvature than the age.  They also show that the introduction of uncertainties somewhat reduces the magnitude of this curvature by generally flattening the LFs.  The magnitude of this effect appears to be relatively insignificant for LFs with uncertainties based on the data for either HSB or LSB  galaxies.\par

While other factors such as the assumed IMF, the star formation history, and blending also determine the shape of the H{\sc ii} region LF, our Monte Carlo simulations provide some insight into the origin(s) of the difference in shapes among the LFs plotted in Fig.\ \ref{loclf} in terms of age and cluster size.  From the simulated LFs shown in Fig.\ \ref{mclfs}, it appears that the power-law shape of the LFs seen throughout LSB galaxy discs and in the most LSB regions of HSB discs are plausibly the result of clusters forming with relatively large values of $N_{\ast , max}$ and that the curvature at the bright-end seen for the HSB regions of HSB discs requires smaller values of $N_{\ast , max}$.  However, the paucity of highly luminous H{\sc ii} regions within the LSB regions of dics (see Fig.\ \ref{loclum}) suggests that it is highly unlikely that $N_{\ast , max}$ is smaller for clusters within HSB regions than for those within LSB regions.  From Fig.\ \ref{mclfs}, it can be seen that if the clusters within LSB regions are relatively old ($\sim$6 Myr) and the clusters within HSB regions are significantly younger ($\!\,^{<}_{\sim}$1 Myr), the value of $N_{\ast , max}$ can be roughly the same.  Since the effect of blending has not been included in the simulated LFs but likely affects the observed LFs, we cannot provide accurate estimates of the typical ages of the clusters within HSB and LSB regions using these simulations.  However, it appears that qualitatively, the simulated LFs suggest that the typical star cluster within LSB regions is significantly older than the typical HSB region cluster.  We note that a similar explanation has been proposed for the difference in LF shapes between arm and inter-arm H{\sc ii} regions within spiral galaxies with arm regions being younger on average \citep{oey98,von90}.\par
%How large depends on the age of the clusters, but it appears that $N_{\ast , max}$ is at least $\sim 50$ for relatively large ages of 6 My, but must me much larger ($\sim 400$) for very young clusters.  The curvature at the bright-end seen for the HSB regions of HSB discs appears to require the opposite, low values of $N_{\ast , max}$ ($\sim 10$) with larger ages ($\sim 6$ Myr) or young ages with moderate values of $N_{\ast , max}$ ($\sim 50$).  As we demonstrated in \S 3.3, blending will affect the shape of the H{\sc ii} region LF, regardless of the value of $\mu_{B,lim}$, but it is likely not the main cause of the difference in shape between the HSB and LSB galaxy LFs.  However, we must note that our Monte Carlo simulations do not include the effects of blending which makes the numbers we have quoted for ages and values of $N_{\ast , max}$ here only rough estimates.\par
As the LFs in Fig.\ \ref{loclf} demonstrate, the higher surface brightness regions of HSB discs tend to have more H{\sc ii} regions per square kiloparsec than LSB galaxies have throughout their discs.  The Monte Carlo-simulated LFs seem to imply that these more densely packed areas tend to have younger clusters on average.  Since the HSB discs have spiral arms more frequently than the LSB discs in our samples, this could point to a fundamental difference between spiral-arm driven star formation and the more sporadic star formation that is likely prominent in LSB discs, particularly for Sm and Im galaxies \citep[see, e.g., ][]{bot84,deb95,hel05}.  The results displayed in Fig.\ \ref{loclf} and \ref{lfshape} would further imply that there may be a transition from spiral-arm driven star formation to a more erratic process within HSB discs that typically occurs at $\mu_{B} \sim 24$.  The galaxy M101 provides some verification of this notion as the luminous H{\sc ii} regions in the outer area of its disc suggest a more erratic process of star formation than is seen within the disc where spiral arms are prevalent \citep[see, e.g.,][]{sco92}.\par
It is also possible that the change in the shape of the H{\sc ii} region LF that appears to occur within HSB galaxies at a local surface brightness of $\mu_{B} \sim 24$ is linked to a change in the cold gas density.  For example, the surface brightness profile we have measured for M83 reaches $\mu_{B} = 24$ at an apparent radius of about 4.4 arcmin.  From the radial profiles presented by \citet{thi05}, the gas surface density at this radius is about 25 M$_{\odot}$ pc$^{-2}$, which is roughly six times larger than the typical critical gas density for star formation determined for similar spiral galaxies \citep[see, e.g.,][]{ken89}.  However, at this radius, the profile for the molecular hydrogen drops quickly, and the slope of $\mbox{log } \Sigma_{\mbox{\scriptsize H}_{2}} (R)$ abruptly becomes steeper by a factor of $\sim 3$.  At this radius, the gas-phase oxygen abundance of M83 has also dropped by factor of about two compared to what it is in the inner parts of the disk \citep{dia84}.  This may be the underlying cause of the lack of both CO gas and the dust grains needed to facilitate the conversion of atomic to molecular hydrogen.  This example demonstrates that it is plausible that the apparent transition to a more sporadic mode of star formation is linked to a change in physical conditions that has caused a sharp decrease in the amount molecular gas that is readily available in the outer regions of spiral galaxy discs.\par
Overall, the results we have presented demonstrate that while local conditions appear to have a large effect on star formation within all galaxy discs, it appears that the change in H{\sc ii} region properties from higher to lower density environments is significantly more pronounced within HSB discs than it is within LSB discs.  The most plausible explanation appears to be that the star formation within the outer parts of HSB discs and throughout LSB discs is more sporadic than that which occurs within the inner regions of HSB discs.  Since LSB discs tend to have spiral arms less frequently, little if any molecular gas, and relatively low metallicities, it may be that this sporadic mode of star formation tends to dominate within regions that have relatively low amounts of dust and molecular gas that are also relatively stable against global instabilities like spiral density waves.  This is also consistent with the physical conditions of the outer regions of spiral galaxy discs \citep[see, e.g.,][]{dia84,thi05}, which points to a common physical origin for the similarly shaped H{\sc ii} region LFs of both LSB discs and the LSB regions of HSB discs.

\section*{Acknowledgements}
The authors would like to thank the referee for useful comments and suggestions.  This research was performed while the lead author held a National Research Council Research Associateship Award at the Naval Research Laboratory.  Basic research in astronomy at the Naval Research Laboratory is supported by 6.1 base funding.

\bsp

\label{lastpage}

\end{document}